\documentstyle[11pt,epsf]{article}
\input epsf
\begin{document}
\baselineskip 100pt
\renewcommand{\baselinestretch}{1.5}
\renewcommand{\arraystretch}{0.666666666}
{\large
\parskip.2in  
\newcommand{\be}{\begin{equation}}
\newcommand{\ee}{\end{equation}}
\newcommand{\br}{\bar}
\newcommand{\fr}{\frac}
\newcommand{\lm}{\lambda}
\newcommand{\ra}{\rightarrow}
\newcommand{\al}{\alpha}
\newcommand{\bt}{\beta}
\newcommand{\pr}{\partial}
\newcommand{\hs}{\hspace{5mm}}
\newcommand{\up}{\upsilon}
\newcommand{\dg}{\dagger}
\newcommand{\ve}{\varepsilon}
\newcommand{\acc}{\\[3mm]}
 
\hfill DTP\,/43-96
 
\vskip 1truein
\bigskip
\begin{center}
{\bf Soliton Dynamics in a Novel Discrete $\rm{O}$(3) Sigma Model in
(2+1) Dimensions.} \end{center}
 
\vskip 1truein
\bigskip
\begin{center}
T{\small HEODORA} I{\small OANNIDOU}\\
{\sl Department of Mathematical Sciences, University of Durham,\\
Durham DH1 3LE, UK}
\end{center}
 
\vskip 2truein
\bigskip  

{\bf Abstract.}
The $\rm{O}$(3) sigma model in two spatial dimensions admits topological 
(Bogomol'nyi) lower bound on its energy.
This paper proposes a lattice version of this system which maintains the 
Bogomol'nyi bound and allows the explicit construction of static 
solitons on the lattice.
Numerical simulations show that these lattice solitons are unstable under 
small perturbations; in fact, their size changes linearly with time.

To appear in Nonlinearity.

\newpage

{\bf {\rm {\bf 1}}. Introduction.}
 
The nonlinear $\rm{O}$(3) sigma model in (2+1) dimensions is a popular
model in theoretical physics; the static system is integrable and of
Bogomol'nyi type (all minimal energy solutions can be obtained by solving
the Bogomol'nyi equations).
As a result, one can explicitly write down soliton solutions of arbitrary 
degree in term of rational functions \cite{BP}; but the model is scale 
invariant and therefore, its solitons have no fixed size and so their 
stability is a central question.
Under small perturbations they shrink towards infinitely tall spikes of 
zero width or may spread out, with this expansion continuing indefinitely.
That this indeed happens is confirmed by numerical experments \cite{LPZ,PZ}.
General time-dependent solutions cannot be constructed explicitly, and so 
it is natural to investigate numerical evolution techniques which 
discretize the partial differential equations.
 
Given a continuum field theory, there are many different lattice systems
which reduce to it in the continuum limit.
In systems where there are topological configurations (instantons,
monopoles, etc) one often has a  Bogomol'nyi bound which is related to
the stability of the topological objects in question.
If the bound is maintained on the lattice, the topological objects will be 
well-behaved even when their size is comparable to the lattice spacing.
Lattice versions of these systems are important for purposes of 
numerical computations but they have, generally, not preserved the 
Bogomol'nyi bound. 
Few years ago, Leese \cite{L} discretized the 
(unmodified) $\rm{O}$(3) sigma model in $(2+1)$ dimensions.
He imposed radial symmetry,  made the radial coordinate $r$
discrete and found a reduced lattice system with  
Bogomol'nyi bound.
But, although the topological lower bound can be attained, the 
minimum-energy configurations are not explicit.
On the other hand, Ward \cite{W} described a lattice version of this 
model with Bogomol'nyi bound, without any symmetry constraint.
In this general case, however, the lower bound cannot be attained.

The purpose of this paper is to present a lattice version of the
$\rm{O}$(3) sigma model in two space dimensions, in which the Bogomol'nyi
bound is maintained.
The primary aim is not to simulate the continuum system, but rather to
define an alternative lattice system with similar properties but more
convenient to study numerically.
Following Leese, only field configurations for which the energy density 
(and not necessarily the fields) is radially symmetric will be considered 
here, so that in effect one obtains a one-dimensional system and therefore,
the construction of the discrete Bogomol'nyi equations is less
complicated.
Solutions of these equations (which were obtained analytically)  are then 
used as the basis for a numerical study of soliton stability.
In fact we study the shrinking of lattice solitons and try to estimate 
analytically the rate of this shrinking for a soliton and 2-soliton 
configuration.
It was shown \cite{L1} that a collision of two solitons can induce their 
shrinking, therefore we can then follow one of these solitons and study 
its behaviour.
However, due to the soliton movement it is difficult to be very 
quantitative about the soliton shrinking.
For this reason it is convenient the 
restriction to axial symmetry.
It is worth remarking that if radial symmetry is imposed on the fields 
\cite{MY} then the model is integrable, and an inverse scattering 
transformation exist.
But this restriction requires that the obtained solutions have  
topological charge zero.

The rest of this paper is arranged as follows.
In the next section we describe initially, the familiar continuum 
$\rm{O}$(3) sigma model in $(2+1)$ dimensions then reparametrize the 
fields in order to impose radial symmetry and, finally, discretize the model.
In section $3$ we study the dynamics of the 2-soliton configuration at 
low shrinking velocities using the slow-motion  approximation, 
make approximate analytic predictions of its behaviour, and compare 
these with numerical results, while in section $4$ we investigate the 
properties of the lattice $\rm{O}(3)$ solitons numerically.
 
{\bf {\rm {\bf 2}}. The Lattice $\rm{O}$(3) Sigma Model.}
 
Let us begin with a brief review of the continuum  $\rm{O}$(3) sigma
model in two space dimensions.
The field \(\mbox{\boldmath $\phi$}\) is a unit 3-vector field on ${\bf
R^2}$ (i.e. a smooth function from ${\bf R^2}$ to the target space ${\bf 
S^2}$), with the boundary condition \(\mbox{\boldmath $\phi$}\) $\ra$
\(\mbox{\boldmath $\phi_0$}\) as $r \ra \infty$ in  ${\bf R^2}$.  
Here \(\mbox{\boldmath$\phi_0$}\) is some fixed point on the image sphere
${\bf S^2}$.
Hence there are distinct topological sectors classified  by an integer 
{\cal $k$} (topological charge), which represents the number of times ${\bf 
R^2}$ is wrapped around ${\bf S^2}$. 
Roughly speaking, {\cal $k$} is the number of solitons.
The potential energy of the field is $E_p=(8\pi)^{-1}\int[(\pr_x
\mbox{\boldmath $\phi$})^2+(\pr_y\mbox{\boldmath $\phi$})^2]\,dx\,dy$ 
and the appropriate Bogomol'nyi argument gives the bound $E_p \geq$   
{\cal $|k|$}.
There are fields which attain this lower bound (such minimum-energy fields
will be called solitons in what follows).
Since $E_p$ is invariant under the scaling transformation
\(\mbox{\boldmath $\phi$}\)$(x^j) \mapsto$ \(\mbox{\boldmath $\phi$}\)$(\lm
x^j)$ these configurations are metastable rather than stable (their size 
is not fixed).

From now on we will restrict attention to fields which are invariant   
under simultaneous rotations and reflections in space and target space.
Thus we assume that $\mbox{\boldmath $\phi$}=(\mbox{\boldmath
$\phi$}^\al, \, \mbox{\boldmath $\phi$}^3)$ with $\al=1,2$ is of the 
hedgehog form 
\be
\begin{array}{llcl}
\mbox{\boldmath $\phi$}^\al=\sin \,g(r, t)\,\mbox{\cal $k$}^\al,\hs \hs
\hs \mbox{\boldmath $\phi$}^3=\cos \,g(r, t),
\label{hed}
\end{array}
\ee
characterized by its topological charge {\cal $k$}, defining the 
unit vector $\mbox{\cal $k$}^\al=(\cos\mbox{\cal $k$}\theta, \,   
\sin\mbox{\cal $k$}\theta)$ in (\ref{hed}) in terms of the azimuthal 
angle $\theta$; and by the real (profile) function $g$ of the polar 
coordinates and $t$ which satisfies certain boundary 
conditions. 
The corresponding potential energy of the field (\ref{hed}) is
\be
E_p=\fr{1}{4} \int_{0}^{\infty} (r\,g'^2+\fr{ \mbox{\cal $k$}^2}{r}
\sin^2g)\,dr,
\label{enec}
\ee
where $g'=dg/dr$.
This is normalized so that a static configuration has {\cal $k$} energy.
The boundary conditions are $g(0, t)=\pi$, in order to ensure a unique 
definition of $\mbox{\boldmath $\phi$}$ at the
origin and $g(r, t) \ra 0$ as $r \ra \infty$, so that $E_p$ converges.

The standard Bogomol'nyi argument \cite{Bog} is
\begin{eqnarray}
0 &\leq& \fr{1}{4}\int_0^\infty (\sqrt{r} g'+\fr{\mbox{\cal 
$k$}}{\sqrt{r}} \sin g)^2\,dr \nonumber \acc
& =& E_p-\fr{\mbox{\cal$k$}}{2} \int_0^\infty  \pr_r(\cos g)\, dr 
\nonumber \acc 
& = &E_p-\mbox{\cal$k$}.
\label{boun}
\end{eqnarray}
So the energy $E_p$ is bounded below by {\cal $k$}; and $E_p$ equals {\cal 
$k$} if and only if $g'=-\mbox{\cal$k$} \sin g/r$, the solution of which 
is the static {\cal $k$}-soliton configuration
\be
g(r)=2\arctan\left(\fr{a}{r^{\mbox{\cal $k$}}}\right),
\label{csol}
\ee
located at the origin, with $a$ being a positive real constant which 
determines the soliton size. 
If $a$ is large, the soliton configuration is flat and broad; 
while if $a$ is small, it is tall but narrow.
In fact, the height of the configuration (maximum of the 
energy density) is proportional to $a^{-2/\mbox{\cal $k$}}$; while its 
radius (width) is proportional to $a^{1/\mbox{\cal $k$}}$.
Notice that, for  $\mbox{\cal $k$}=0$ the field is constant and the energy 
density is zero
everywhere; while for  $\mbox{\cal $k$}=1$ the configuration looks like a
lump peaked  at the origin; and for  $\mbox{\cal $k$} > 1$ it is a ring
centered at the origin.
In what follows, we will assume that in all cases $\mbox{\cal $k$} > 0$,
since taking $\mbox{\cal $k$}=0$ does not test the ability of the model to
handle nontrivial topologies.

So far all we have done is to re-express the {\cal $k$}-soliton solution 
in terms of a real field $g$, which is a function of the polar radius $r$.
It will now been seen how this description is useful in constructing 
discrete analogues of the Bogomol'nyi equations.

From now on, $r$ becomes a discrete variable, with lattice spacing $h$.
So the real-valued field $g(r, t)$ depends on the continuous variable
$t$, and the discrete variable $r=nh$ ($n \in {\bf Z}$, $n \ge 0$).
The subscript $+$ denotes forward shift, i.e. $g_+(r, t)=g(r+h, t)=g((n+1) 
h, t)$; and so the forward difference is given by $\Delta g=(g_+-g)/h$.
The question of how best to incorporate topological ideas into a lattice 
formulation has been the subject of much discussion, especially in 
lattice gauge theories contexts (cf. \cite{G}).
One approach, following Speight and Ward \cite{SW}, is to  begin with  
the same function $\cos g$ as appears in (\ref{boun}) and reconstruct the 
inequality, i.e. 
\be
\mbox{\cal $k$} (\Delta \cos g)=-D_n F_n,
\ee
where $D_n \ra \sqrt{r} g'$ and $F_n \ra \mbox{\cal $k$} \sin g / 
\sqrt{r}$ in the  continuum limit $h \ra 0$.
The formula  $\mbox{\cal $k$} (\Delta \cos g)=-2 \mbox{\cal 
$k$}/h\sin(\fr{g_+-g}{2}) \sin(\fr{g_++g}{2})$ suggests the 
choices
\begin{eqnarray}
D_n&=&\fr{2 f(h) \sqrt{n}}{\sqrt{h}} \sin
\left(\fr{g_+-g}{2}\right),\nonumber \acc
F_n&=&\fr{\mbox{\cal $k$}}{f(h) \sqrt{h n}} \sin
\left(\fr{g_++g}{2}\right), \hs \hs n > 0,
\label{lsol}  
\end{eqnarray}
where $f(h)$ is an arbitrary function of the lattice spacing with 
constraints $f(h) \ra 1$ as $h \ra 0$ and $f(h) > \sqrt{\mbox{\cal
$k$}/2}$ (see below).
The origin must be treated in a special way since (\ref{lsol}) are
undefined when $n=0$.
One possibility is to arrange that $D_0+F_0=0$ identically.
So choose
\be
D_0\equiv-F_0=\sqrt{\fr{2\mbox{\cal $k$}}{h}} \cos \left(\fr{g(h,
t)}{2}\right).
\ee

The potential energy of the lattice $\rm{O}$(3) sigma model 
field is defined to be
\begin{eqnarray}
E_p&=&\fr{h}{4} \sum_{n=0}^\infty \left(D_n^2+F_n^2\right)\nonumber\\
&=&\mbox{\cal $k$} \cos^2 \left(\fr{g(h, t)}{2}\right)+\sum_{n=1}^\infty
\,[\,f^2\, n \sin^2
\left(\fr{g_+-g}{2}\right)+\fr{\mbox{\cal $k$}^2}{4 f^2\, n} \sin^2
\left(\fr{g_++g}{2}\right)].
\label{ene}
\end{eqnarray}
As in the continuum case, it follows that $E_p$ is bounded below by {\cal 
$k$}; and the  minimum is attained if and only if $D_n+F_n=0$.
For models with different lattice spacing the effect of 
$f(h)$ is to decrease the importance of the $\sin^2 g$ term in the energy 
density, although the total energy is still the same as in the continuum 
({\cal $k$} in our units).

The kinetic energy can be defined by the simple choice
\be
E_k=\fr{h^2}{4}\sum_{n=1}^\infty n\, \dot{g}^2,\acc
\ee
where $\dot{g}=dg/dt$.
The boundary condition on $g$ is that it should tend to zero at spatial 
infinity; this guarantees finite energy.
For such fields, the total energy $E_t=E_p+E_k$ is bounded below by {\cal 
$k$} ; and this lower bound is attained if and only if $\dot{g}=0$, and
$D_n+F_n=0$ for $n > 0$.
[Recall that $D_0+F_0=0$ identically.]

This latter condition, i.e. $D_n+F_n=0$, is called the 
Bogomol'nyi equation.
It is a first-order difference equation, whose solutions (for the 
aforementioned boundary conditions) minimize the potential energy, and 
therefore, are also static solutions of the Euler-Lagrange 
equations 
\be
\sum_{n=1}^\infty n\, \ddot{g}(n h, t)=-\fr{2}{h^2} \fr{\pr{E_p}}{\pr g},\\
\label{leq}
\ee
since $\pr{E_p} / \pr g =0$ at a minimum.
So using the discrete Bogomol'nyi equations one gets first-order 
equations whose solutions are also static solutions of the second-order 
equations of motion.
Moreover, these solutions have energy which is at its topological minimum 
value.

The Bogomol'nyi equation $D_n+F_n=0$, may also be written as
\be
\tan \fr{g_{+}}{2}=\fr{2 f^2 \,n-\mbox{\cal $k$}}{2 f^2\, n+\mbox{\cal $k$}} 
\tan\fr{g}{2}, \hs \hs n > 0,
\label{boge}
\ee
from which one sees that the function $f(h)$ should be greater than 
$\sqrt{\mbox{\cal $k$}/2}$ in order the profile function to be monotonic.
So for ${\mbox{\cal $k$}}=1,2$ the form $f(h)=1+h$ will work for all $h$;
while for ${\mbox{\cal $k$}} \ge 3$ we need a condition on $h$, i.e. $h > 
\sqrt{{\mbox{\cal $k$}}/2}-1$.

The solution of (\ref{boge}) can be written down explicitly, i.e.
\be
g(n h)=\Bigg\{ \begin{array}{llcl}
\pi, \hs \hs \hs \hs&  n=0,\acc
2 \arctan(z_1 {\cal Z}_n),\hs &  n > 0,
\end{array}
\label{soll}
\ee
where
\be
{\cal Z}_n=\fr{\Gamma(n-\mbox{\cal $k$}/2f^2)\,\Gamma(1+\mbox{\cal 
$k$}/2f^2)}{\Gamma(n+\mbox{\cal $k$}/2f^2)\,\Gamma(1-\mbox{\cal 
$k$}/2f^2)}, 
\ee	
and $z_1$  is an arbitrary positive constant which specifies, as in the 
continuum model, the soliton size.
This is a static lattice {\cal $k$}-soliton solution located at the 
origin; which corresponds to a minimum of the energy in the {\cal $k$} 
sector and thus, it is stable under perturbations which remain in that 
sector.
From now on, we will concentrate on the $k=1,2$ sectors  and 
therefore, we will take on the first one the function $f(h)$ to be 
constant and equal to unity, for any $h$; while on the second one to be 
$f(h)=1+h$, for small $h$.

It would be nice to have a lattice analogue of the configuration width 
$a^{1/k}$, which appeared in (\ref{csol}).
One possibility is to set
\be
a_n=(n h)^{\mbox{\cal $k$}}  \tan \fr{g(n h)}{2},
\ee
and then to define $a=\lim_{n \ra \infty} a_n$, provided this limit exists.
Indeed, $a$  is proportional to $z_1 \,h^{\mbox{\cal 
$k$}}$.
If $E_p=\sum_{n=0}^\infty E_{p_{n}}$, the energy density at the origin
is $E_{p_{0}}=\mbox{\cal$k$}/(1+z_1^2)$; 
therefore, $E_{p_{0}}$ is close to the Bogomol'nyi bound as $z_1 \ra 0$. 
A diagram illustrating the profiles of the function $g(n h)$ and the 
energy densities profiles are represented in figure 1 for $\mbox{\cal 
$k$}=1$ and $\mbox{\cal $k$}=2$ with $z_1=15$ and $h=0.19$.

The situation we wish to study is that of an isolated perturbed static 
{\cal $k$}-soliton configuration and investigate the effects of 
the perturbation.
As it costs them no energy to shrink or expand they can shrink to 
almost a zero width (radius) configuration in the energy density plot.
Since the soliton configuration is described by a few points on a lattice it 
is difficult to decide what is meant by its width and how to calculate it.
The lattice analogue will be a field configuration with 
$g(0, t)=\pi$ (due to the boundary conditions) and $g(n h,t)=0$, for $n > 0$.
In fact, this corresponds to a spike soliton (of almost zero width) in 
the continuum and thus, our interest lies in the study of the time 
dependence of the shrinking of a $\mbox{\cal $k$}$-soliton configuration.

Since, there is no explicit solution in this case, one has to resort to
approximation, or to numerical solutions of the equations of motion
(\ref{leq}), namely 
\begin{eqnarray}
\ddot{g}&=&\fr{1}{h^2}[\, \mbox{\cal $k$}
\sin g(h, t)+f^2
\sin (g(2h, t)-g(h, t))]-\fr{\mbox{\cal $k$}^2}{4 f^2\, h^2} \sin
(g(2h, t)+g(h, t)),
\hs \,\,\,\,\,\,  n=1 \nonumber, \acc
n \ddot{g}&=&\fr{f^2}{h^2}[n \sin (g_+ -g)-(n-1) \sin
(g-g_{-})]-\fr{\mbox{\cal $k$}^2}{4 f^2\, h^2}[\fr{\sin
(g_+ +g)}{n}+\fr{\sin(g+g_-)}{n-1}],  \hs  n > 1. \nonumber \\
\label{leq1}
\end{eqnarray}

\begin{figure}[b]
\unitlength1cm
\begin{picture}(5,6)
\put(0,8){(a)}
\put(3,3.5){$\mbox{\cal $k$}=2$}
\put(4.25,4.5){$\mbox{\cal $k$}=1$}
\put(0,5){$g(n h)$}
\put(7,0.75){$n$}
\epsfxsize=1.25cm
\epsffile{prof.ps}
\end{picture}
\par
\par
\par
\hfill
\begin{picture}(15.75,10)
\put(0,8){(b)}
\put(2,2.5){$\mbox{\cal $k$}=1$}
\put(3,8){$\mbox{\cal $k$}=2$}
\put(0,5.5){$\fr{E_{p_n}}{2 \pi n h}$}
\put(7,0.75){$n$}
\epsfxsize=1.25cm
\epsffile{ener.ps}
\end{picture} 
\par
\hfill
\caption{(a) Profiles of $g$ for charges $\mbox{\cal $k$}=1,2$.
(b) Profiles of the energy densities $E_{p_{n}}/(2 \pi n h)$ for the
charges in (a). The $\mbox{\cal $k$}=2$ energy density is ring shaped.}
\end{figure}

{\bf {\rm {\bf 3}}. The Slow-Motion Approximation.}

There is a fundamental difference between the cases $\mbox{\cal $k$}=1$ and 
$\mbox{\cal $k$}> 1$, which becomes apparent when one considers the 
slow-motion  approximation, originally proposed in 
connection with monopole scattering \cite{M}.
In this scheme one assumes that the field $g$ is a static solution, but 
slightly perturbed.
More precisely, since the energy is conserved, and due to the existence of 
the Bogomol'nyi bound, we may assume that a {\cal $k$}-soliton dynamics 
is obtained by restricting $g$ to have the form of (\ref{soll}), with 
$z_1$ now becoming a dynamical variable $z_1(t)$.
So the number of degrees of freedom is reduced from infinite to one.
These static solutions form a manifold, which is equipped with a natural 
metric coming from the kinetic energy, and the evolution is given by the 
resulting geodesics.
Since every configuration of the form (\ref{soll}) has the same potential 
energy, the kinetic energy may be taken as the Lagrangian; thus, the 
corresponding Euler-Lagrange equations are precisely the geodesic 
equations associated with the aforementioned metric.
This approximation is a good one if the speeds are small (if $E_k$ 
is small compared to $E_p=\mbox{{\cal $k$}}$). 

For $\mbox{\cal $k$}=1$, the requirement of finite kinetic energy means that 
$z_1$ should be independent of $t$ at spatial infinity, so ruling out the 
slow-motion approximation.
In other words, taking $z_1$ to be a function only of $t$ leads to a 
divergent kinetic energy.
But when $\mbox{\cal $k$} > 1$ there are sufficient powers of $n$ in the 
denominator of (\ref{soll}) to keep the energy finite.
For this case the slow-motion approximation has been considered in 
\cite{L1,W1} in order to  study the dynamics of {\bf CP$^1$} 
lumps. 
However, they have not looked at the speed of the shrinking in any detailed.
Let us concentrate on the  $\mbox{\cal $k$}= 2$ topological 
sector, where two solitons are sitting on top of each other at the 
origin, forming a ring structure.
The Lagrangian is
\begin{eqnarray}
L&=&E_k-E_p  \nonumber \\
&=&l(z_1)\,\dot{z}_1^2-2,
\end{eqnarray}
where
\be
l(z_1)=h^2 \sum_{n=1}^\infty \fr{n {\cal Z}_n^2}{(1+z_1^2 {\cal Z}_n^2)^2}.
\label{lfu}
\ee

The Euler-Lagrange equation of the system is
\be
2 \,l(z_1)\, \ddot{z}_1+l'(z_1) \,\dot{z}_1^2=0 \hs \hs \Rightarrow \hs \hs 
\fr{d}{dt}\left(l(z_1)\, \dot{z}_1^2\right)=0.
\label{eq}
\ee
which may be reduced to quadratures:
\begin{eqnarray}
\upsilon t &=& \int_{c}^{z_1(t)} 
\sqrt{\fr{l(\tilde{z}_1)}{l(c)}}\,d\tilde{z}_1\nonumber \acc
&\equiv & \Lambda_h (z_1),
\end{eqnarray}
where $z_1(0)=c$, $\dot{z}_1(0)=\upsilon$.
Recall that,  $z_1$ determines the configuration size which evolves with 
$t$; more precisely,  $\sqrt{c}h$  is the initial width of the configuration 
and, 
$\upsilon$ is the initial rate of  change of the configuration width in 
each lattice site per unit time.
In fact, $\upsilon < 0$ corresponds to an initial contraction and 
$\upsilon > 0$ to an initial expansion.

The function $\Lambda_h (z_1)$ decreasing or increasing depending on the 
value of $z_1(t)$, which corresponds to contraction or expansion of the 
configuration. 
It is easily inverted to give the time variation 
of the configuration size, i.e. 
\be
z_1(t)=\Lambda^{-1}_h (\upsilon t).
\label{rat}
\ee
Recall that, the lattice analogue of the configuration width is 
proportional to $\sqrt{z_1}h$.
In fact, the time taken for the configuration to shrink from
the initial width  to zero, i.e. to become a spike, is
\be
t_c=\fr{\Lambda_h(0)}{\upsilon}.
\ee
We believe that this analytical approximation is accurate for small 
$|\upsilon|$.

The accuracy of the approximation has been tested numerically using a 
fully-explicit fourth-order Runge-Kutta algorithm with fixed time step 
0.0053. 
The initial condition was a static 2-soliton profile whose width 
we perturbed to shrink with initial velocity $0.1$ lattice site 
per unit time ($\upsilon=-0.1 h$). 
Simulations of duration 2985 time units were performed for $h=0.01$.
Inspection of the rate of change of ${z}_1(t)=\tan(g(h,t)/2)$ reveals close 
agreement with $\dot{z}_1(t)$ calculated from (\ref{rat}) (see figure 2).

\begin{figure}[b]
\hspace{.5in}
\unitlength1cm
\begin{picture}(4,12)
\put(-1,5.25){$|\dot{z}_1|$}
\put(7,-.25){$t$}
\epsfxsize=14cm
\epsffile{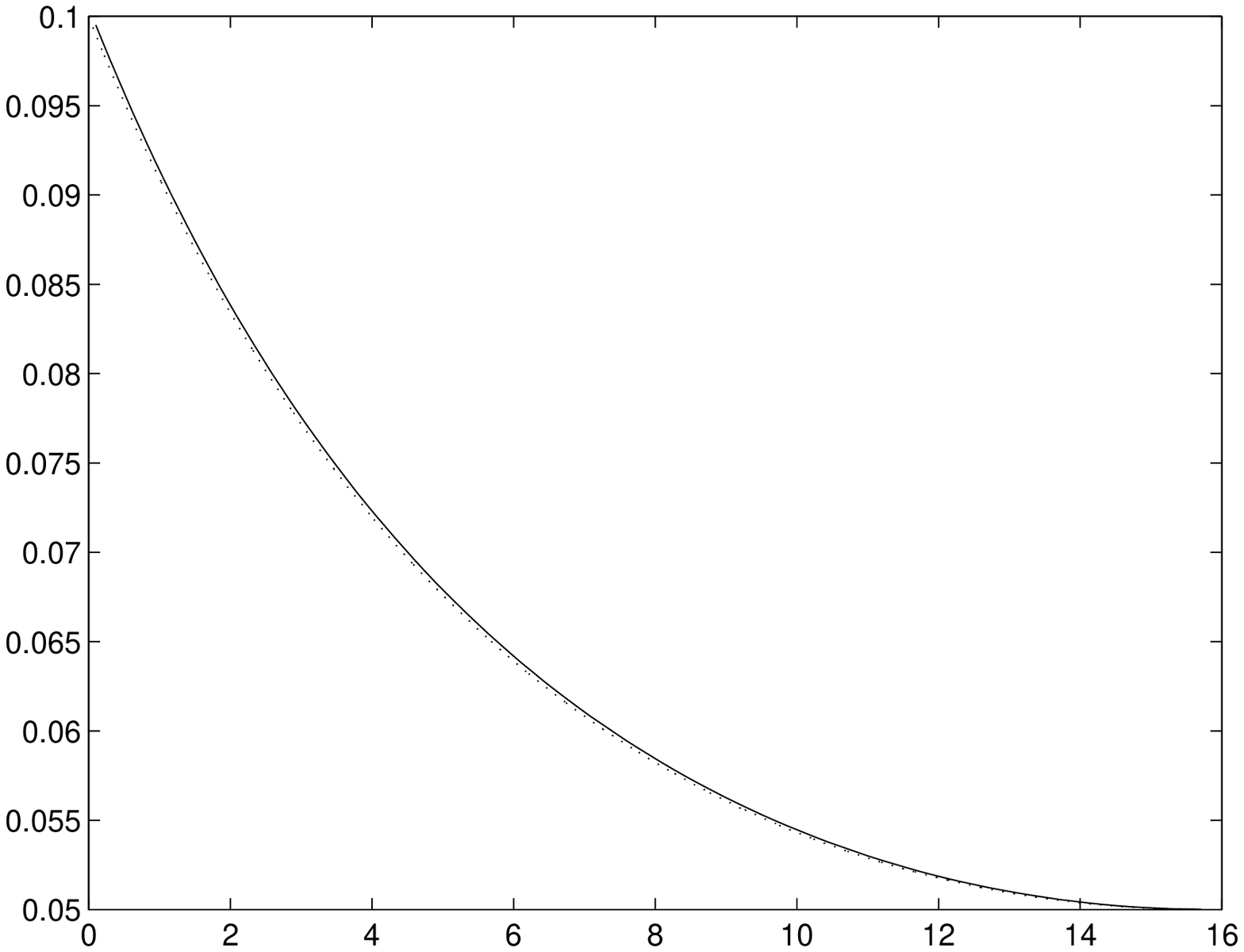}
\end{picture}
\par
\hfill
\caption{The time variation of $|\dot{z_1}|$ in the analytic slow-motion approximation (solid line),
and also for the numerical evolution (dotted line).}
\end{figure}

{\bf {\rm {\bf 4}}. Dynamics of the Lattice $\rm{O}$(3) Solitons.}

Since the slow-motion approximation is expected to fail at high velocities 
(except for small $h$), we incorporate the notion of lattice solitons in a 
full numerical evolution scheme.
Throughout the simulations, the extensive use of the difference 
equations (\ref{leq1}) have not revealed any instabilities (the total 
energy is conserved).

The lattice formulation necessarily has a spatial boundary at $n=n_{max}$, 
say. 
Hence, the quantities that we are going to use in order to 
study the soliton dynamics  will be calculated within some radius 
($n_{max}$). 
Moreover, the infinite sums on the energies will be truncated.
In fact, for $\mbox{\cal $k$}=1$ the finiteness of the grid imposes an 
artificial cutoff which provides a finiteness in the energy.
On the boundary though, the fields are taken to be fixed in time, i.e.
\be
g((n_{max}+1)\, h, t)=g((n_{max}+1)\, h,0),
\ee
since if one attempts to apply boundary conditions which allow the field 
to change with time at arbitrary  distances, then the total energy of 
the system for $\mbox{\cal $k$}=1$, grows rapidly and without bound.
One may also, choose  absorbing boundary conditions or may place the 
boundary far enough from the configuration (i.e., no radiation effects).
But, in this scheme, the choice of the boundary conditions has no impact 
on the rate of shrinking.

Moving on to the question of initial data, there are clearly many different 
types of perturbation which we could apply to the configuration, the only 
restriction being that we do not perturb the field close to the boundary.
Since the evolution equations (\ref{leq1}) are second order the initial data 
must specify the field values $g(n h, t)$ and its time derivatives 
$\dot{g}(n h, t)$ at $t=0$.
So the field configuration at $t=0$ is taken to be the static lattice 
one (\ref{soll}) but slightly perturbed, i.e. 
\be
\dot{g}(n h, t)\left|\right._{t=0}=\frac{2 \upsilon {\cal Z}_n}{1+z_1^2 
{\cal Z}_n^2}.\acc
\ee
Physically the picture is this:
there is a continuous interpolation between the inner region where 
$\upsilon$ is the amplitude of the perturbation (as in the 
slow-motion approximation) and the outer one where there is no 
perturbation at all. 
This class of perturbation reveals all the qualitative types of behaviour 
that can occur.

So we have a {\cal $k$}-soliton configuration whose centre remains fixed, 
but whose radius decreases to a minimum (close to zero) and then 
increases again.
More precisely, the initial perturbation (for $\upsilon < 0$) tends to 
shrink the  configuration, while large burst of radiation travel outwards 
at the  speed of light (see figure 3), together with a residual motion in 
the central region occupied by the soliton.
When the radiation reaches the boundary, is reflected back, reabsorbed by 
the configuration which expands and then another pulse is emitted a short 
time later; and so the process repeats.
The data were produced by the aforementioned Runge-Kutta algorithm for 
$\upsilon=-0.1$ and $z_1=1$, on a lattice of unit spacing ($h=1$) in 
the $\mbox{\cal $k$}=1$ sector.

We are interested in the speed of the shrinking in detail.
In order to analyze the results of the numerical simulations 
we look at the dynamical quantity $g(h, t)$.
Since we are on a lattice, the soliton will be highly localized (a 
spike) when the profile function at the first site ($n=1$) and 
consequently, at all others ($n>1$) will be zero.
Then, the soliton configuration occupies essentially only one 
lattice site, while the slope of the curve $g(h,t)$ correspond to the 
power law of shrinking.

In figure 4 we present the time dependence of the field $g(h, t)$ 
for a single soliton and a 2-soliton configuration.
The results are derived from a relatively small mesh $n_{max}=200$ (in 
fact, they do not change for larger mesh sizes), for $\upsilon=-0.1$ and 
$z_1=1$.
In the single soliton case, we make the simple choice $f(h)=1$ whereas 
$h=1$; while in the 2-soliton case, we take $f(h)=1+h$ with $h=0.01$.
(Note that, figure 4(b) corresponds to figure 2.)
The field configuration saturates the Bogomol'nyi bound throughout the 
numerical evolution and due to that in the $\mbox{\cal $k$}=1$ sector 
the  lattice spacing is comparable to the soliton size without 
compromising its behaviour.
As it is clear for the graphs the curves are nearly straight, confirming 
the power law for the rate of shrinking.
The linear curve in figure 4(b) is due to the fact that
the lattice spacing is small  compare to the size of the topological soliton 
and thus, the model is closer to the continuum one.
Let us conclude with the observation that our results are consistent with 
the ones obtained by studying the continuum ${\rm O}(3)$ sigma model 
(cf. \cite{PZ}).

\begin{figure}[b]
\hspace{.5in}
\unitlength1cm
\begin{picture}(4,12)
\put(-2.1,6){$g(n h, t)|_{t=11}$}
\put(7,-.25){$n$}
\epsfxsize=14cm
\epsffile{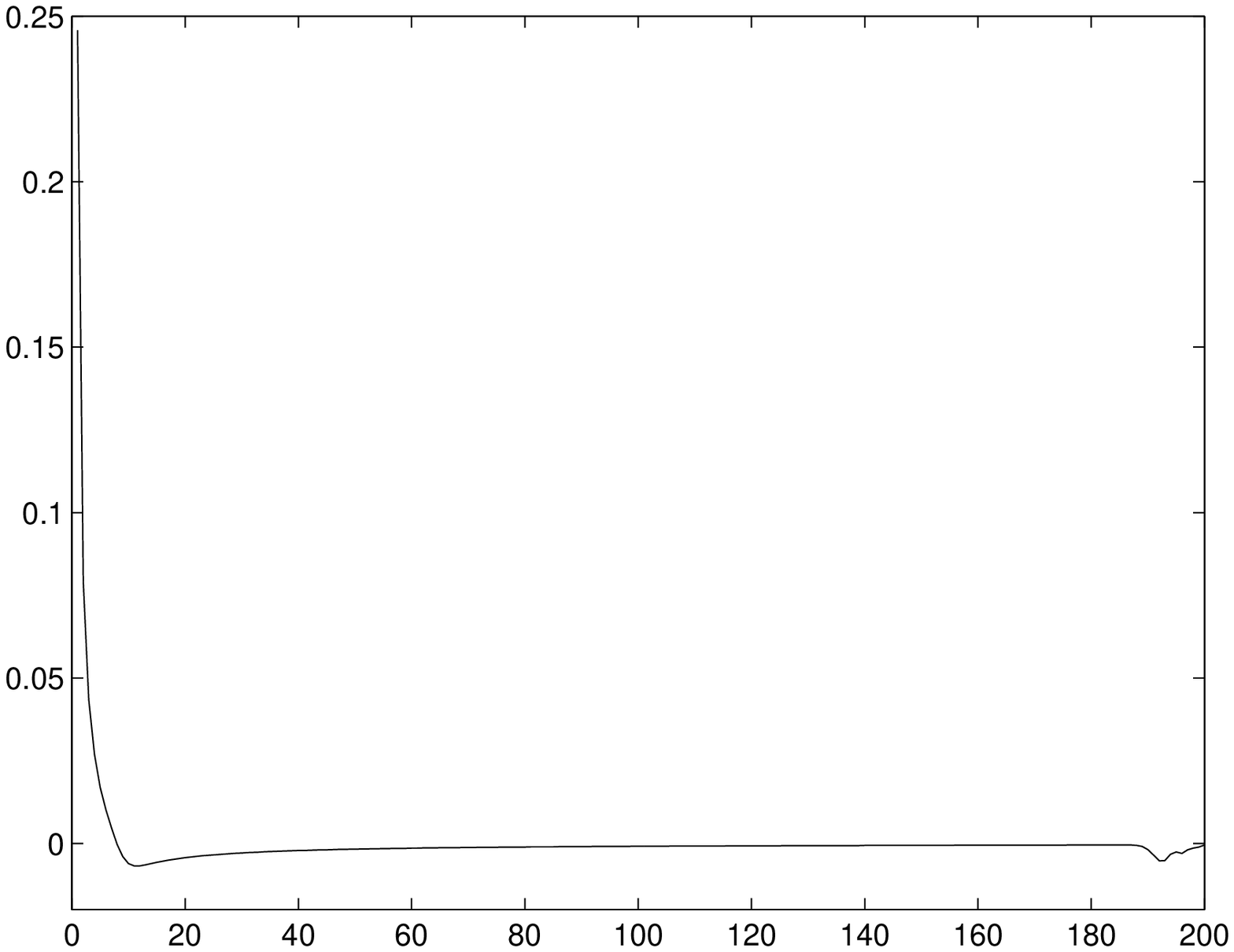}
\end{picture}
\par
\hfill
\caption{Radiation emitted by the 1-soliton solution.}
\end{figure}

\begin{figure}[b]
\hspace{0.45in}
\unitlength1cm
\begin{picture}(3.5,10)
\put(-1,8){(a)}
\put(-1.2,5){$g(h, t)$}
\put(6,-.5){$t$}
\epsfxsize=12cm
\epsffile{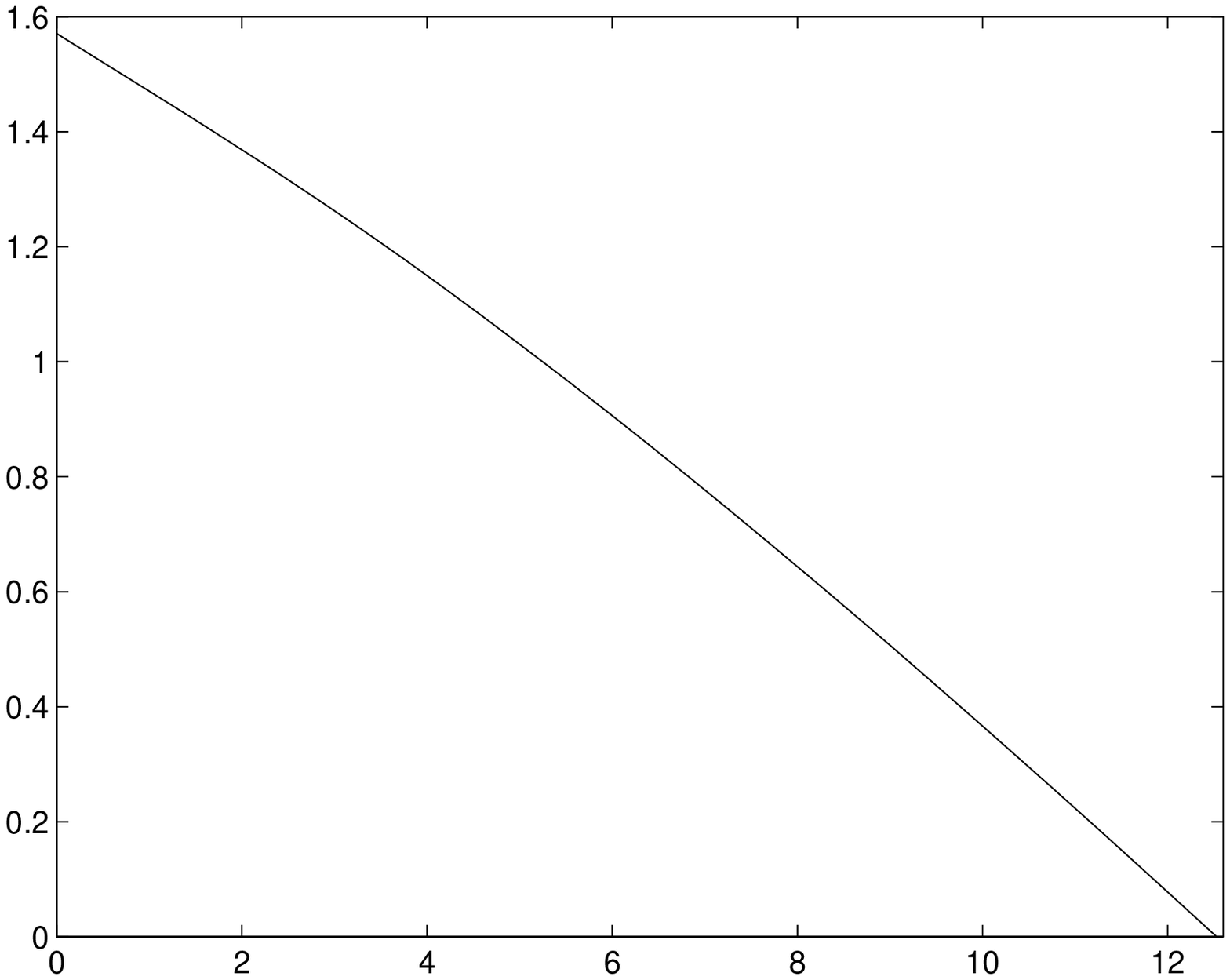}
\end{picture}
\par
\par
\par
\hfill
\begin{picture}(14.65,12)
\put(-1,8){(b)}
\put(-1.25,5){$g(h, t)$}
\put(6,-.5){$t$}
\epsfxsize=12cm
\epsffile{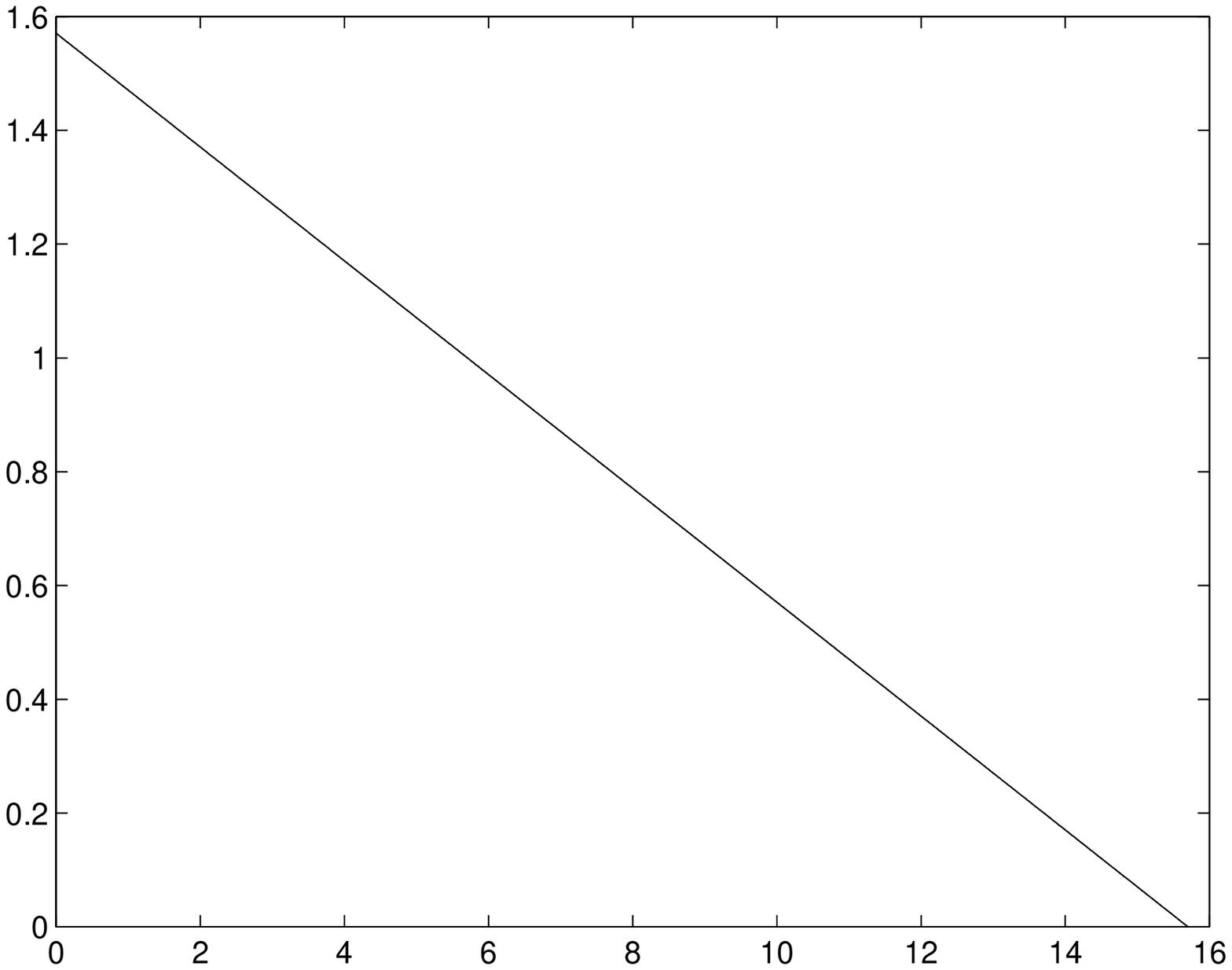}
\end{picture}
\hfill
\caption{The variation of $g(h, t)$ over the range (a) $0 \leq t \leq
12.6$ for a slowly shrinking 1-soliton lump and (b) $0 \leq t \leq 16$ for a
slowly shrinking 2-soliton ring, for the numerical evolution.}
\end{figure}

{\bf {\rm {\bf 5}}. Conclusions.}

We have studied the time evolution of the lattice $\rm{O}$(3) sigma 
solitons which were allowed to shrink.
We have used two different methods to analyze numerically the soliton 
shrinking.
The first method is based on the slow-motion approximation and leads to an
ordinary differential equation.
The second one consists of integrating numerically the 
(1+1)-dimensional semi-discrete equations.
From all the numerical studies we have performed we conclude that a 
single soliton and a 2-soliton configuration of the $\rm{O}$(3) lattice 
sigma model does shrink to a highly localized (spike) lattice soliton,  
linearly with time.
If one is close to the continuum limit, in a sense that the lattice 
spacing is small compared to the size of the topological solitons, then 
there may not be much difference between various lattice versions of the 
continuum system.
An advantage of the lattice model described in this paper is that the 
lattice spacing can be relatively large without compromising the soliton 
dynamics.

Let us conclude by stressing once again that the slow-motion 
approximation works  very well.

{\bf Acknowledgments.}

I would like to thank Richard Ward and Wojtek Zakrzewski for helpful 
discussions. This research was supported by EC ERBFMBICT950035.

\vfill\eject

\newpage

\end{document}